\input epsf
\input amssym 
\catcode`@=11                                   
\catcode`\|=12                                  
\catcode`\&=4                                   

\newcount\ncols         \ncols=\z@              
\newcount\nrows         \nrows=\z@              
\newcount\curcol        \curcol=\z@             
     
\newdimen\thinsize      \thinsize=0.6pt         
\newdimen\thicksize     \thicksize=1.5pt        

\newif\iftableinfo      \tableinfotrue          
\newif\ifcentertables   \centertablestrue       
%
%
     
\let\plaincr=\cr                        
\let\plainspan=\span                    
\let\plaintab=&                         
\let\lparen=(                           
\let\NX=\noexpand                       

     
\def\ruledtable{\relax                          
    \@BeginRuledTable                           
    \@RuledTable}


\def\@BeginRuledTable{
   \ncols=0\nrows=0                             
   \begingroup                                  
    \offinterlineskip                           
    \def~{\phantom{0}}
    \def\span{\plainspan\omit\relax\colcount\plainspan}
    \let\cr=\crrule                             
    \let\CR=\crthick                            
    \let\nr=\crnorule                           
    \let\|=\Vb                                  
%
%
    \ifx\tablestrut\undefined\relax             
    \else\let\tstrut=\tablestrut\fi             
    \catcode`\|=13 \catcode`\&=13\relax         
    \TableActive                                
    \curcol=1                                   
%
%
    \ifdim\tablewidth>-\maxdimen\relax          %
      \edef\@Halign{\NX\halign to \NX\tablewidth\NX\bgroup\TablePreamble}%
      \tabskip=0pt plus 1fil                    
    \else                                       %
      \edef\@Halign{\NX\halign\NX\bgroup\TablePreamble}%
      \tabskip=0pt                              
    \fi                                         %
%
%
    \ifcentertables                             
       \ifhmode\vskip 0pt\fi                    
       \line\bgroup\hss                         
    \else\hbox\bgroup                           
    \fi}


\long\def\@RuledTable#1\endruledtable{
   \vrule width\thicksize                       
     \vbox{\@Halign                             
       \thickrule                               
       #1\relax                                 
       \tstrut                                  
       \plaincr\thickrule                       
     \egroup}
   \vrule width\thicksize                       
   \ifcentertables\hss\fi\egroup                
  \endgroup                                     
  \global\tablewidth=-\maxdimen                 
  \iftableinfo                                  
      \immediate\write16{[Nrows=\the\nrows, Ncols=\the\ncols]}%
   \fi}
     

\def\TablePreamble{
   \linecount                           
   \TableItem{####}
   \plaintab\plaintab                   
   \TableItem{####}
   \plaincr}


\def\@TableItem#1{
   \hfil\tablespace                             
   #1\relax                                     
   \tablespace\hfil                             
    }%

\def\@tableright#1{
   \hfil\tablespace\relax               
   #1\relax                             
   \tablespace\relax}

\def\@tableleft#1{
   \tablespace\relax                    
   #1\relax                             
   \tablespace\hfil}

\let\TableItem=\@TableItem              
     
\def\RightJustifyTables{\let\TableItem=\@tableright}
\def\LeftJustifyTables{\let\TableItem=\@tableleft}
\def\NoJustifyTables{\let\TableItem=\@TableItem}

\def\LooseTables{\let\tablespace=\quad}
\def\TightTables{\let\tablespace=\space}
\LooseTables                                    

%

\newdimen\tablewidth    \tablewidth=-\maxdimen  


\def\setRuledStrut{
   \dimen@=\baselineskip                        
   \advance\dimen@ by-\normalbaselineskip       
   \ifdim\dimen@<.5ex \dimen@=.5ex\fi           
   \setbox0=\hbox{\lparen}
   \dimen1=\dimen@ \advance\dimen1 by \ht0      
   \dimen2=\dimen@ \advance\dimen2 by \dp0      
   \def\tstrut{\vrule height\dimen1 depth\dimen2 width\z@}%
   }%

\def\tstrut{\vrule height 3.1ex depth 1.2ex width 0pt}


\def\bigitem#1{
   \setbox0=\hbox{#1}
   \dimen1 =\ht0 \dimen2 =\dp0                  
   \dimen@ =\baselines@ve                       
   \advance\dimen@ by-\normalbaselineskip       
   \ifdim\dimen@<.25ex \dimen@=.25ex\fi         
   \advance\dimen1 by \dimen@                   
   \advance\dimen2 by \dimen@                   
   \vrule height\dimen1 depth\dimen2 width\z@   
   \copy0}

     
%

     
\def\nextcolumn#1{
   \plaintab\omit#1\relax\colcount              
   \plaintab}
     
\def\tab{
   \nextcolumn{\relax}}


\def\vb{
   \nextcolumn{\vrule width\thinsize}}

\def\Vb{
   \nextcolumn{\vrule width\thicksize}}


     
{\catcode`\|=13 \let|0
 \catcode`\&=13 \let&0
 \gdef\TableActive{\let|=\vb \let&=\tab}%
}


\def\crrule{\relax                      
   \tstrut                              
   \plaincr\tablerule                   
  }%

\def\crthick{\relax                     
   \tstrut                              
   \plaincr\thickrule                   
  }%
     
\def\crnorule{\relax                    
   \tstrut                              
   \plaincr                             
   }%
   

     
\def\tablerule{\noalign{\hrule height\thinsize depth 0pt}}%
\def\thickrule{\noalign{\hrule height\thicksize depth 0pt}}%


%
%
%
     

\def\linecount{\relax\global\ncols=\curcol      
   \global\curcol=1                             
   \global\advance\nrows by 1\relax}
     
\def\colcount{\relax                            %
   \global\advance\curcol by 1\relax}


\newdimen\parasize      \parasize=4in           

%

%

\def\begintable{\relax                          
    \@BeginRuledTable                           
    \@begintable}

\long\def\@begintable#1\endtable{
   \@RuledTable#1\endruledtable}


\catcode`@=12                                   


\catcode`\@11\relax
\newif\ify@autoscale \y@autoscaletrue \def\Yautoscale#1{\ifnum #1=0
  \y@autoscalefalse\else\y@autoscaletrue\fi}
\newdimen\y@b@xdim
\newdimen\y@boxdim \y@boxdim=13pt
\def\Yboxdim#1{\y@autoscalefalse\y@boxdim=#1}
\newdimen\y@linethick    \y@linethick=.3pt
\def\Ylinethick#1{\y@linethick=#1}
\newskip\y@interspace \y@interspace=0ex plus 0.3ex
\def\Yinterspace#1{\y@interspace=#1}
\newif\ify@vcenter   \y@vcenterfalse
\def\Yvcentermath#1{\ifnum #1=0 \y@vcenterfalse\else\y@vcentertrue\fi}
\newif\ify@stdtext   \y@stdtextfalse
\def\Ystdtext#1{\ifnum #1=0 \y@stdtextfalse\else\y@stdtexttrue\fi}
\newif\ify@enable@skew   \y@enable@skewfalse
\expandafter\ifx\csname enableskew\endcsname\relax
 \y@enable@skewfalse \else \y@enable@skewtrue\fi
\def\y@vr{\vrule height0.8\y@b@xdim width\y@linethick depth 0.2\y@b@xdim}
\def\y@emptybox{\y@vr\hbox to \y@b@xdim{\hfil}}
\ify@enable@skew
 \def\y@abcbox#1{\if :#1\else
   \y@vr\hbox to \y@b@xdim{\hfil#1\hfil}\fi}
 \def\y@mathabcbox#1{\if :#1\else
   \y@vr\hbox to \y@b@xdim{\hfil$#1$\hfil}\fi}
\else
 \def\y@abcbox#1{\y@vr\hbox to \y@b@xdim{\hfil#1\hfil}}
 \def\y@mathabcbox#1{\y@vr\hbox to \y@b@xdim{\hfil$#1$\hfil}}
\fi
\def\y@setdim{%
  \ify@autoscale%
   \ifvoid1\else\typeout{Package youngtab: box1 not free! Expect an
     error!}\fi%
   \setbox1=\hbox{A}\y@b@xdim=1.6\ht1 \setbox1=\hbox{}\box1%
  \else\y@b@xdim=\y@boxdim \advance\y@b@xdim by -2\y@linethick
  \fi}
\newcount\y@counter
\newif\ify@islastarg
\def\y@lastargtest#1,#2 {\if\space #2 \y@islastargtrue
  \else\y@islastargfalse\fi}
\def\y@emptyboxes#1{\y@counter=#1\loop\ifnum\y@counter>0
  \advance\y@counter by -1 \y@emptybox\repeat}
\def\y@nelineemptyboxes#1{%
  \vbox{%
    \hrule height\y@linethick%
    \hbox{\y@emptyboxes{#1}\y@vr}
    \hrule height\y@linethick}\vskip-\y@linethick}
\def\yng(#1){%
  \y@setdim%
  \hskip\y@interspace%
  \ifmmode\ify@vcenter\vcenter\fi\fi{%
  \y@lastargtest#1,
  \vbox{\offinterlineskip
    \ify@islastarg
     \y@nelineemptyboxes{#1}
    \else
     \y@ungempty(#1)
    \fi}}\hskip\y@interspace}
\def\y@ungempty(#1,#2){%
  \y@nelineemptyboxes{#1}
  \y@lastargtest#2,
  \ify@islastarg
   \y@nelineemptyboxes{#2}
  \else
   \y@ungempty(#2)
  \fi}
\def\y@nelettertest#1#2. {\if\space #2 \y@islastargtrue
  \else\y@islastargfalse\fi}
\def\y@abcboxes#1#2.{%
  \ify@stdtext\y@abcbox#1\else\y@mathabcbox#1\fi%
  \y@nelettertest #2.
  \ify@islastarg\unskip%
   \ify@stdtext\y@abcbox{#2}\else\y@mathabcbox{#2}\fi%
  \else\y@abcboxes#2.\fi}
 \newdimen\y@full@b@xdim
 \newcount\y@m@veright@cnt
\ify@enable@skew
 \def\y@get@m@veright@cnt#1#2.{%
   \if :#1 \advance\y@m@veright@cnt by 1\y@get@m@veright@cnt#2.\fi}
 \let\y@setdim@=\y@setdim
 \def\y@setdim{%
   \y@setdim@ \y@full@b@xdim=\y@b@xdim
   \advance\y@full@b@xdim by 1\y@linethick}
 \def\y@m@veright@ifskew#1{
   \y@m@veright@cnt=0 \y@get@m@veright@cnt#1.
   \moveright \y@m@veright@cnt\y@full@b@xdim}
\else
 \def\y@m@veright@ifskew#1{}
\fi
\def\y@nelineabcboxes#1{%
  \y@nelettertest #1.
  \ify@islastarg
   \y@m@veright@ifskew{#1}
    \vbox{
      \hrule height\y@linethick%
      \hbox{\ify@stdtext\y@abcbox#1\else\y@mathabcbox#1\fi\y@vr}
      \hrule height\y@linethick}\vskip-\y@linethick
  \else
   \y@m@veright@ifskew{#1}
    \vbox{
      \hrule height\y@linethick%
      \hbox{\y@abcboxes #1.\y@vr}%
      \hrule height\y@linethick}\vskip-\y@linethick
  \fi}
\def\young(#1){%
  \y@setdim%
  \hskip\y@interspace%
  \y@lastargtest#1,
  \ifmmode\ify@vcenter\vcenter\fi\fi{%
  \vbox{\offinterlineskip
    \ify@islastarg\y@nelineabcboxes{#1}%
    \else\y@ungabc(#1)%
    \fi}}\hskip\y@interspace}
\def\y@ungabc(#1,#2){%
  \y@nelineabcboxes{#1}%
  \y@lastargtest#2,
  \ify@islastarg\y@nelineabcboxes{#2}%
  \else\y@ungabc(#2)%
  \fi}
\catcode`\@12\relax
 


\newfam\scrfam
\batchmode\font\tenscr=rsfs10 \errorstopmode
\ifx\tenscr\nullfont
        \message{rsfs script font not available. Replacing with calligraphic.}
        
\else   
        \font\sevenscr=rsfs7
        \font\fivescr=rsfs5
        \skewchar\tenscr='177 \skewchar\sevenscr='177 \skewchar\fivescr='177
        \textfont\scrfam=\tenscr \scriptfont\scrfam=\sevenscr
        \scriptscriptfont\scrfam=\fivescr

\fi
\catcode`\@=11
\newfam\frakfam
\batchmode\font\tenfrak=eufm10 \errorstopmode
\ifx\tenfrak\nullfont
        \message{eufm font not available. Replacing with italic.}
        
\else
    
    \font\sevenfrak=eufm7 \font\fivefrak=eufm5
    \textfont\frakfam=\tenfrak
    \scriptfont\frakfam=\sevenfrak \scriptscriptfont\frakfam=\fivefrak
    
\fi
\catcode`\@=\active
\newfam\msbfam
\batchmode\font\twelvemsb=msbm10 scaled\magstep1 \errorstopmode
\ifx\twelvemsb\nullfont\def\Bbb{\bf}

    \message{Blackboard bold not available. Replacing with boldface.}
\else   \catcode`\@=11
        \font\tenmsb=msbm10 \font\sevenmsb=msbm7 \font\fivemsb=msbm5
        \textfont\msbfam=\tenmsb
        \scriptfont\msbfam=\sevenmsb \scriptscriptfont\msbfam=\fivemsb
        \def\Bbb{\relax\expandafter\Bbb@}
        \def\Bbb@#1{{\Bbb@@{#1}}}
        \def\Bbb@@#1{\fam\msbfam\relax#1}
        \catcode`\@=\active

\fi
\newfam\cpfam
\def\sectionfonts{\relax
    \textfont0=\twelvecp          \scriptfont0=\ninecp
      \scriptscriptfont0=\sixrm
    \textfont1=\twelvei           \scriptfont1=\ninei
      \scriptscriptfont1=\sixi
    \textfont2=\twelvesy           \scriptfont2=\ninesy
      \scriptscriptfont2=\sixsy
    \textfont3=\twelveex          \scriptfont3=\tenex
      \scriptscriptfont3=\tenex
    \textfont\itfam=\twelveit     \scriptfont\itfam=\nineit
    \textfont\slfam=\twelvesl     \scriptfont\slfam=\ninesl
    \textfont\bffam=\twelvebf     \scriptfont\bffam=\ninebf
      \scriptscriptfont\bffam=\sixbf
    \textfont\ttfam=\twelvett
    \textfont\cpfam=\twelvecp
}
        \font\eightrm=cmr8              \def\xrm{\eightrm}
        \font\eightbf=cmbx8             \def\xbf{\eightbf}
        \font\eightit=cmti10 at 8pt     \def\xit{\eightit}
                       
        \font\sixrm=cmr6                
        \font\eighttt=cmtt8             
        
        \font\eighti=cmmi8              \def\xold{\eighti}
        \font\eightib=cmmib8             \def\xbold{\eightib}
        \font\teni=cmmi10               \def\old{\teni}
        \font\ninei=cmmi9
        \font\tencp=cmcsc10
        \font\ninecp=cmcsc9

        \font\twelvei=cmmi12
        \font\twelvecp=cmcsc10 scaled\magstep1
        
        \font\fiverm=cmr5
        
        \font\twelvesy=cmsy12
        \font\ninesy=cmsy9
        \font\sixsy=cmsy6
        \font\twelveex=cmex12

        \font\twelveit=cmti12
        \font\nineit=cmti9
        
        \font\twelvesl=cmsl12
        \font\ninesl=cmsl9
        
        \font\twelvebf=cmbx12
        \font\ninebf=cmbx9
        \font\sixbf=cmbx6
        \font\twelvett=cmtt12

        \font\sixi=cmmi6

\batchmode\font\tenhelvbold=phvb at10pt \errorstopmode
\ifx\tenhelvbold\nullfont
        \message{phvb font not available. Replacing with cmr.}
    \font\tenhelvbold=cmb10   
    
    \font\fourteenhelvbold=cmb14
    
  \else
    \font\tenhelvbold=phvb at10pt   
     at12pt
    \font\fourteenhelvbold=phvb at14pt
     at16pt
\fi

\def\noblackbox{\overfullrule=0pt}
\noblackbox

\font\eightmi=cmmi8
\font\sixmi=cmmi6
\font\fivemi=cmmi5

\font\eightsy=cmsy8
\font\sixsy=cmsy6
\font\fivesy=cmsy5

\font\eightsl=cmsl8

\def\eightpoint{
\def\rm{\fam0\eightrm}
\textfont0=\eightrm \scriptfont0=\sixrm \scriptscriptfont0=\fiverm
\textfont1=\eightmi  \scriptfont1=\sixmi  \scriptscriptfont1=\fivemi
\textfont2=\eightsy \scriptfont2=\sixsy \scriptscriptfont2=\fivesy
\textfont3=\tenex   \scriptfont3=\tenex \scriptscriptfont3=\tenex
\textfont\itfam=\eightit \def\it{\fam\itfam\eightit}
\textfont\slfam=\eightsl \def\sl{\fam\slfam\eightsl}
\textfont\ttfam=\eighttt \def\tt{\fam\ttfam\eighttt}
\textfont\bffam=\eightbf \scriptfont\bffam=\sixbf 
                         \scriptscriptfont\bffam=\fivebf
                         \def\bf{\fam\bffam\eightbf}
\normalbaselineskip=10pt}

\newtoks\headtext
\headline={\hfill}
\def\makeheadline{\vbox to 0pt{\vss\noindent\the\headline\break
\hbox to\hsize{\hfill}}
        \vskip2\baselineskip}
\newcount\infootnote
\infootnote=0
\def\foot#1#2{\infootnote=1
\footnote{${}^{#1}$}{\vtop{\baselineskip=.75\baselineskip
\advance\hsize by
-\parindent{\eightpoint\rm\hskip-\parindent #2}\hfill\vskip\parskip}}\infootnote=0$\,$}
\newcount\refcount
\refcount=1
\newwrite\refwrite
\def\oldsize{\ifnum\infootnote=1\xold\else\old\fi}
\def\ref#1#2{
    \def#1{{{\oldsize\the\refcount}}\ifnum\the\refcount=1\immediate\openout\refwrite=\jobname.refs\fi\immediate\write\refwrite{\item{[{\xold\the\refcount}]}
    #2\hfill\par\vskip-2pt}\xdef#1{{\noexpand\oldsize\the\refcount}}\global\advance\refcount by 1}
    }
\def\refout{\catcode`\@=11
        \xrm\immediate\closeout\refwrite
        \vskip2\baselineskip
        {\noindent\tenhelvbold References}\hfill
        \par\nobreak\vskip\baselineskip
        \baselineskip=.75\baselineskip
        \input\jobname.refs
        \baselineskip=4\baselineskip \divide\baselineskip by 3
        \catcode`\@=\active\rm}

\def\hepth#1{\href{http://arxiv.org/abs/hep-th/#1}{arXiv:hep-th/{\xold#1}}}

\def\arxiv#1#2{\href{http://arxiv.org/abs/#1.#2}{arXiv:{\xold#1}.{\xold#2}}}
\def\jhep#1#2#3#4{\href{http://jhep.sissa.it/stdsearch?paper=#2\%28#3\%29#4}{J. High Energy Phys. {\xbold #1#2} ({\xold#3}) {\xold#4}}}

\def\CQG#1#2#3{Class. Quantum Grav. {\xbold#1} ({\xold#2}) {\xold#3}}

\def\JHEP{\jhep}

\def\LMP#1#2#3{Lett. Math. Phys. {\xbold#1} ({\xold#2}) {\xold#3}}
\def\MPLA#1#2#3{Mod. Phys. Lett. {\xbf A}{\xbold#1} ({\xold#2}) {\xold#3}}
\def\NPB#1#2#3{Nucl. Phys. {\xbf B}{\xbold#1} ({\xold#2}) {\xold#3}}

\def\PLB#1#2#3{Phys. Lett. {\xbf B}{\xbold#1} ({\xold#2}) {\xold#3}}

\newcount\sectioncount
\sectioncount=0
\def\section#1#2{\global\eqcount=0
    \global\subsectioncount=0
        \global\advance\sectioncount by 1
    \ifnum\sectioncount>1
            \vskip2\baselineskip
    \fi
    \noindent
       \line{\sectionfonts\twelvecp\the\sectioncount. #2\hfill}
        \par\nobreak\vskip.8\baselineskip\noindent
        \xdef#1{{\old\the\sectioncount}}}
\newcount\subsectioncount
\def\subsection#1#2{\global\advance\subsectioncount by 1
    \par\nobreak\vskip.8\baselineskip\noindent
    \line{\tencp\the\sectioncount.\the\subsectioncount. #2\hfill}
    \vskip.5\baselineskip\noindent
    \xdef#1{{\old\the\sectioncount}.{\old\the\subsectioncount}}}
\newcount\appendixcount
\appendixcount=0
\def\appendix#1{\global\eqcount=0
        \global\advance\appendixcount by 1
        \vskip2\baselineskip\noindent
        \ifnum\the\appendixcount=1
        \hbox{\twelvecp Appendix A: #1\hfill}
        \par\nobreak\vskip\baselineskip\noindent\fi
    \ifnum\the\appendixcount=2
        \hbox{\twelvecp Appendix B: #1\hfill}
        \par\nobreak\vskip\baselineskip\noindent\fi
    \ifnum\the\appendixcount=3
        \hbox{\twelvecp Appendix C: #1\hfill}
        \par\nobreak\vskip\baselineskip\noindent\fi}

\newcount\eqcount
\eqcount=0
\def\Eqn#1{\global\advance\eqcount by 1
\ifnum\the\sectioncount=0
    \xdef#1{{\old\the\eqcount}}
    \eqno({\oldstyle\the\eqcount})
\else
        \ifnum\the\appendixcount=0
            \xdef#1{{\old\the\sectioncount}.{\old\the\eqcount}}
                \eqno({\oldstyle\the\sectioncount}.{\oldstyle\the\eqcount})\fi
        \ifnum\the\appendixcount=1
            \xdef#1{{\oldstyle A}.{\old\the\eqcount}}
                \eqno({\oldstyle A}.{\oldstyle\the\eqcount})\fi
        \ifnum\the\appendixcount=2
            \xdef#1{{\oldstyle B}.{\old\the\eqcount}}
                \eqno({\oldstyle B}.{\oldstyle\the\eqcount})\fi
        \ifnum\the\appendixcount=3
            \xdef#1{{\oldstyle C}.{\old\the\eqcount}}
                \eqno({\oldstyle C}.{\oldstyle\the\eqcount})\fi
\fi}
\def\eqn{\global\advance\eqcount by 1
\ifnum\the\sectioncount=0
    \eqno({\oldstyle\the\eqcount})
\else
        \ifnum\the\appendixcount=0
                \eqno({\oldstyle\the\sectioncount}.{\oldstyle\the\eqcount})\fi
        \ifnum\the\appendixcount=1
                \eqno({\oldstyle A}.{\oldstyle\the\eqcount})\fi
        \ifnum\the\appendixcount=2
                \eqno({\oldstyle B}.{\oldstyle\the\eqcount})\fi
        \ifnum\the\appendixcount=3
                \eqno({\oldstyle C}.{\oldstyle\the\eqcount})\fi
\fi}
\def\multi{\global\advance\eqcount by 1}
\def\multieq#1#2{
    \ifnum\the\sectioncount=0
        \eqno({\oldstyle\the\eqcount})
         \xdef#1{{\old\the\eqcount#2}}
    \else
        \xdef#1{{\old\the\sectioncount}.{\old\the\eqcount}#2}
        \eqno{({\oldstyle\the\sectioncount}.{\oldstyle\the\eqcount}#2)}
    \fi}

\newtoks\url
\def\Href#1#2{\catcode`\#=12\url={#1}\catcode`\#=\active#2}
\def\href#1#2{{#2}}

\parskip=3.5pt plus .3pt minus .3pt
\baselineskip=14pt plus .1pt minus .05pt
\lineskip=.5pt plus .05pt minus .05pt
\lineskiplimit=.5pt
\abovedisplayskip=18pt plus 4pt minus 2pt
\belowdisplayskip=\abovedisplayskip
\hsize=14cm
\vsize=20.8cm
\hoffset=1.5cm
\voffset=1.5cm
\frenchspacing
\footline={}
\raggedbottom

\def\ss{\scriptstyle}

\def\*{\partial}
\def\punkt{\,\,.}
\def\komma{\,\,,}

\def\={\!=\!}
\def\small#1{{\hbox{$#1$}}}

\def\fraction#1{\small{1\over#1}}
\def\fr{\fraction}
\def\Fraction#1#2{\small{#1\over#2}}
\def\Fr{\Fraction}

\def\eg{{\tenit e.g.}}

\def\ie{{\tenit i.e.}}

\def\a{\alpha}
\def\b{\beta}
\def\c{\gamma}
\def\d{\delta}

\def\g{\gamma}

\def\ra{\rightarrow}

\def\Tr{\hbox{Tr}\,}


\def\Tr{\hbox{Tr}}




\def\l{\lambda}

\def\lra{\longrightarrow}
\def\ra{\rightarrow}

\def\arrowunder#1{\raise4pt\vtop{\baselineskip=0pt\lineskip=0pt
      \ialign{\hfill##\hfill\cr${\ss #1}$\cr$\lra$\cr}}}

\def\<{{<}}
\def\>{{>}}

\def\Q{Q}

\def\lra{\longrightarrow}

\def\ra{\rightarrow}
\def\la{\leftarrow}

\def\rarrowover#1{\vtop{\baselineskip=0pt\lineskip=0pt
      \ialign{\hfill##\hfill\cr$\ra$\cr$#1$\cr}}}

\def\larrowover#1{\vtop{\baselineskip=0pt\lineskip=0pt
      \ialign{\hfill##\hfill\cr$\la$\cr$#1$\cr}}}

\def\<{{<}}
\def\>{{>}}

\def\TableItem#1{
   \hfil\tablespace                             
   #1\relax                                     
   \tablespace                             
    }%
\thicksize=\thinsize

\def\lb{\bar\lambda}
\def\TPsi{\tilde\Psi}

%
%

\ref\CederwallNilssonTsimpisI{M. Cederwall, B.E.W. Nilsson and D. Tsimpis,
{\xit ``The structure of maximally supersymmetric super-Yang--Mills
theory---constraining higher order corrections''},
\jhep{01}{06}{2001}{034} 
[\hepth{0102009}].}

\ref\CederwallNilssonTsimpisII{M. Cederwall, B.E.W. Nilsson and D. Tsimpis,
{\xit ``D=10 super-Yang--Mills at $\ss O(\a'^2)$''},
\JHEP{01}{07}{2001}{042} [\hepth{0104236}].}

\ref\BerkovitsParticle{N. Berkovits, {\xit ``Covariant quantization of
the superparticle using pure spinors''}, \jhep{01}{09}{2001}{016}
[\hepth{0105050}].}

\ref\SpinorialCohomology{M. Cederwall, B.E.W. Nilsson and D. Tsimpis,
{\xit ``Spinorial cohomology and maximally supersymmetric theories''},
\jhep{02}{02}{2002}{009} [\hepth{0110069}];
M. Cederwall, {\xit ``Superspace methods in string theory, supergravity and gauge theory''}, Lectures at the XXXVII Winter School in Theoretical Physics ``New Developments in Fundamental Interactions Theories'',  Karpacz, Poland,  Feb. 6-15, 2001, \hepth{0105176}.}

\ref\Movshev{M. Movshev and A. Schwarz, {\xit ``On maximally
supersymmetric Yang--Mills theories''}, \NPB{681}{2004}{324}
[\hepth{0311132}].}

\ref\BerkovitsI{N. Berkovits,
{\xit ``Super-Poincar\'e covariant quantization of the superstring''},
\jhep{00}{04}{2000}{018} [\hepth{0001035}].}

\ref\BerkovitsNonMinimal{N. Berkovits,
{\xit ``Pure spinor formalism as an N=2 topological string''},
\jhep{05}{10}{2005}{089} [\hepth{0509120}].}

\ref\CederwallNilssonSix{M. Cederwall and B.E.W. Nilsson, {\xit ``Pure
spinors and D=6 super-Yang--Mills''}, \arxiv{0801}{1428}.}

\ref\CGNN{M. Cederwall, U. Gran, M. Nielsen and B.E.W. Nilsson,
{\xit ``Manifestly supersymmetric M-theory''},
\JHEP{00}{10}{2000}{041} [\hepth{0007035}];
{\xit ``Generalised 11-dimensional supergravity''}, \hepth{0010042}.
}

\ref\CGNT{M. Cederwall, U. Gran, B.E.W. Nilsson and D. Tsimpis,
{\xit ``Supersymmetric corrections to eleven-dimen\-sional supergravity''},
\jhep{05}{05}{2005}{052} [\hepth{0409107}].}

\ref\HoweTsimpis{P.S. Howe and D. Tsimpis, {\xit ``On higher order
corrections in M theory''}, \jhep{03}{09}{2003}{038} [\hepth{0305129}].}

\ref\NilssonPure{B.E.W.~Nilsson,
{\xit ``Pure spinors as auxiliary fields in the ten-dimensional
supersymmetric Yang--Mills theory''},
\CQG3{1986}{{\xrm L}41}.}

\ref\HowePureI{P.S. Howe, {\xit ``Pure spinor lines in superspace and
ten-dimensional supersymmetric theories''}, \PLB{258}{1991}{141}.}

\ref\HowePureII{P.S. Howe, {\xit ``Pure spinors, function superspaces
and supergravity theories in ten and eleven dimensions''},
\PLB{273}{1991}{90}.} 

\ref\FreGrassi{P. Fr\'e and P.A. Grassi, {\xit ``Pure spinor formalism
for OSp(N$\ss |$4) backgrounds''}, \arxiv{0807}{0044}.}

\ref\CederwallBLG{M. Cederwall, {\xit ``N=8 superfield formulation of
the Bagger--Lambert--Gustavsson model''}, \jhep{08}{09}{2008}{116} 
[\arxiv{0808}{3242}].}

\ref\CederwallABJM{M. Cederwall, {\xit ``Superfield actions for N=8 and N=6 conformal theories in three dimensions''}, \jhep{08}{10}{2008}{070} 
[\arxiv{0809}{0318}].}

\ref\MarneliusOgren{R. Marnelius and M. \"Ogren, {\xit ``Symmetric
inner products for physical states in BRST quantization''},
\NPB{351}{1991}{474}.} 

\ref\BerkovitsICTP{N. Berkovits, {\xit ``ICTP lectures on covariant
quantization of the superstring''}, proceedings of the ICTP Spring
School on Superstrings and Related Matters, Trieste, Italy, 2002
[\hepth{0209059}.]}

\ref\BerkovitsNekrasovCharacter{N. Berkovits and N. Nekrasov, {\xit
``The character of pure spinors''}, \LMP{74}{2005}{75} [\hepth{0503075}].}
 
\ref\BerkovitsNekrasovReg{N. Berkovits and N. Nekrasov, {\xit
``Multiloop superstring amplitudes from non-minimal pure spinor
formalism''}, 
\jhep{06}{12}{2006}{029} [\hepth{0609012}].}

\ref\GrassiVanhove{P.A. Grassi and P. Vanhove, {\xit ``Higher-loop
amplitudes in the non-minimal pure spinor formalism''},
\jhep{09}{05}{2009}{089} [\arxiv{0903}{3903}].} 

\ref\ElevenSG{E. Cremmer, B. Julia and J. Scherk, 
{\xit ``Supergravity theory in eleven-dimensions''},
\PLB{76}{1978}{409}.}

\ref\ElevenSSSG{L. Brink and P. Howe, 
{\xit ``Eleven-dimensional supergravity on the mass-shell in superspace''},
\PLB{91}{1980}{384};
E. Cremmer and S. Ferrara,
{\xit ``Formulation of eleven-dimensional supergravity in superspace''},
\PLB{91}{1980}{61}.}

\ref\PureSG{M. Cederwall, {\xit ``Towards a manifestly supersymmetric
    action for D=11 supergravity''}, \jhep{10}{01}{2010}{117}
    [\arxiv{0912}{1814}],  
{\xit ``D=11 supergravity with manifest supersymmetry''},
    \MPLA{25}{2010}{3201} [\arxiv{1001}{0112}].}

\ref\ConvConstr{S.J.~Gates, K.S.~Stelle and P.C.~West,
{\xit ``Algebraic origins of superspace constraints in supergravity''},
\NPB{169}{1980}{347}; 
S.J. Gates and W. Siegel, 
{\xit ``Understanding constraints in superspace formulation of supergravity''},
\NPB{163}{1980}{519}.}

\ref\CGNN{M. Cederwall, U. Gran, M. Nielsen and B.E.W. Nilsson, 
{\xit ``Manifestly supersymmetric M-theory''}, 
\JHEP{00}{10}{2000}{041} [\hepth{0007035}];
{\xit ``Generalised 11-dimensional supergravity''}, \hepth{0010042}.}

\ref\BatalinVilkovisky{I.A. Batalin and G.I. Vilkovisky, {\xit ``Gauge
algebra and quantization''}, \PLB{102}{1981}{27}.}

\ref\FusterBVReview{A. Fuster, M. Henneaux and A. Maas, {\xit
``BRST-antifield quantization: a short review''}, \hepth{0506098}.}

\ref\AisakaBerkovits{Y. Aisaka and N. Berkovits, {\xit ``Pure spinor
vertex operators in Siegel gauge and loop amplitude
regularization''}, \jhep{09}{07}{2009}{062} [\arxiv{0903}{3443}].}

\ref\ABC{Y. Aisaka, N.Berkovits and M. Cederwall, unpublished.}


\headtext={}


\hsize=14cm

\vskip8\parskip

\noindent{\fourteenhelvbold
From supergeometry to pure spinors} 

%

\vskip6\parskip

{\tenhelvbold
Martin Cederwall}

\vskip2\parskip

\vbox{\baselineskip=.8\baselineskip
{\eightrm Fundamental Physics}\hfill\break
\null\hskip\parindent{\eightrm Chalmers University of Technology}\hfill\break
\null\hskip\parindent{\eightrm SE 412 96 G\"oteborg, Sweden}\hfill\break
 \catcode`\@=11
\null\hskip\parindent{\eightrm martin.cederwall@chalmers.se}
\catcode`\@=\active
}

\vskip6\parskip

\vbox{\baselineskip=.8\baselineskip\narrower\noindent {\tenhelvbold Abstract:}
 
\noindent{\eightrm 
In this talk, we review how the superspace formulation of maximally
supersymmetric field theories (including supergravity) 
naturally leads to introduction of pure
spinors and pure spinor superfields, and why the formalism provides
off-shell formulations. This approach to pure spinor superfields thus
stresses field-theoretic aspects rather than the first-quantised ones normally
used {\eightit e.g.} in superstring theory.
We discuss how the BRST operator arises and the principles behind
constructions of actions, as well as the general Batalin--Vilkovisky framework.
{\eightit D =} 11 supergravity 
and its recently constructed supersymmetric action [\PureSG] 
is taken as an example throughout the talk.
This is the written version of a lecture given at the 6th Mathematical
Physics Meeting, Belgrade, September 2010.}
\smallskip}




\vskip6\parskip

\noindent Maximally supersymmetric models\foot\star{This means 8
real supersymmetries for scalar multiplets, 16 for vector/tensor
multiplets and 32 for supergravity multiplets}have fields that come in
on-shell supermultiplets. The supersymmetry algebra on the component
fields close (together with gauge transformations) only modulo
equations of motion. In a traditional superfield formulation,
this is a problem, since it implies that supersymmetry can not be
manifested in an action formulation.

For some time, it has been known that the introduction of pure spinors
can solve this problem. In fact, it is turned into an advantage. Such
a formulation does not contradict any no-go theorems against the
existence of auxiliary fields, since the number of component fields
added by the introduction of more bosonic variables is infinite.
In this talk, I will review the quite natural transition from a
traditional superspace formulation of a maximally supersymmetric model
to a formalism with pure spinors, and also discuss some formal
developments. The discussion will, apart from some final remarks,
concern classical field theory, even if one of the eventual goals will
be to examine quantum properties of the models in question, with as
much symmetry as possible manifest. Some aspects will be touched on
only briefly, and in case I am not able to convey the method in a
convincing way, more information can be found in the references.

There is a close relation between supermultiplets and pure
spinors. The algebra of covariant fermionic derivatives in flat
superspace is generically of the form
$$
\{D_\a,D_\b\}=-T_{\a\b}{}^cD_c=-2\g_{\a\b}^cD_c\punkt\Eqn\Torsion
$$
If a bosonic spinor $\l^\a$ is {\it pure}, \ie, if the vector part
$(\l\g^a\l)$ of the spinor bilinear vanishes, the operator
$$
\Q=\l^\a D_\a\Eqn\BRSToperator
$$ 
becomes nilpotent, and may be used as a BRST operator.
This is, schematically, the starting point for pure spinor
superfields. (The details of course depend on the actual space-time and
the amount of supersymmetry. The pure spinor constraint may need to be
further specified. Eq. (\Torsion) may also contain more terms, due to
super-torsion and curvature.) The cohomology of $\Q$ will consist of
supermultiplets, which in case of maximal supersymmetry are
on-shell. The idea of manifesting maximal supersymmetry off-shell by using pure
spinor superfields $\Psi(x,\theta,\l)$ is to find an action whose
equations of motion is $\Q\Psi=0$.

The fact that pure spinors had a r\^ole to play in maximally
supersymmetric models was recognised early by Nilsson [\NilssonPure]
and Howe [\HowePureI,\HowePureII].
Pure spinor superfields were developed with the purpose of covariant
quantisation of superstrings by Berkovits
[\BerkovitsI,\BerkovitsParticle,\BerkovitsICTP,\BerkovitsNonMinimal] and the
cohomological structure was independently discovered in supersymmetric
field theory and supergravity, originally in the context of
higher-derivative deformations
[\CederwallNilssonTsimpisI,\CederwallNilssonTsimpisII,\SpinorialCohomology,\Movshev,\CederwallNilssonSix,\CGNN,\CGNT,\HoweTsimpis].
The present lecture only deals with pure spinors for maximally
supersymmetric field theory.

The canonical example taken to illustrate the mechanisms at play is
$D=10$ super-Yang--Mills theory. I this lecture, I will take the
opportunity to use a supergravity theory, $D=11$ supergravity [\ElevenSG]
as the example. In a sense, this is the only model that fits our
requirements. If we look for a maximal supergravity, the choice is
between this model, and type IIB supergravity in $D=10$, or their
dimensional reductions. Type IIB contains a self-dual tensor field,
which prevents a Lagrangian formulation. So, the choice is $D=11$
supergravity; there is no ``toy model''. 
The situation is a bit more technically 
complicated than for $D=10$ SYM, but I hope you will bear with
this. The structure turns out to be very rewarding.

The component fields of $D=11$ supergravity are
$$
\matrix{
\hbox{metric}\hfill&\qquad g_{mn}\hfill&\qquad\hbox{(bosonic)}\hfill\cr
\hbox{3-form}\hfill&\qquad C_{mnp}\hfill&\qquad\hbox{(bosonic)}\hfill\cr
\hbox{gravitino}\hfill&\qquad \psi_m^\a\hfill&\qquad\hbox{(fermionic)}\hfill\cr
}
$$
The component action takes the form
$$
\eqalign{
S&=\fr{2\kappa^2}\int
d^{11}x\sqrt{-g}\left(R-\fr{48}H^{mnpq}H_{mnpq}\right)\cr
&+\fr{12\kappa^2}\int C\wedge H\wedge H
+\hbox{terms with fermions}\komma\cr}\eqn
$$
where $H=dC$ is the 4-form field strength.

The superspace formulation of $D=11$ supergravity is well known
[\ElevenSSSG]. It follows the standard procedure for supergravity in
superspace. The coordinates $x^m$ are complemented by fermionic
coordinates $\theta^\mu$, and we write $Z^M=(x^m,\theta^\mu)$.
The vielbein (frame) 1-form is extended to a 1-form on superspace with
a flat tangent index:
$$
E^A=dZ^ME_M^A\komma\eqn
$$
$A=(a,\a)$ being the flat index. 
The spin connection 1-form $\Omega_A{}^B$ is Lorentz valued. 
One also defines torsion and
curvature 2-forms 
$$
\eqalign{
T^A&=DE^A=dE^A+E^B\wedge\Omega_B{}^A\komma\cr
R_A{}^B&=d\Omega_A{}^B+\Omega_A{}^C\wedge\Omega_C{}^B\komma
}\eqn
$$ 
which leads to the Bianchi identies
$$
\eqalign{
DT^A&=E^B\wedge R_B{}^A\komma\cr 
DR_A{}^B&=0\punkt\cr
}\eqn
$$
In Einstein (bosonic) gravity, torsion is set to zero. This does not
happen here, as we will see shortly. 

Remember that all components of the vielbein and spin connection are
superfields. We have much too many fields. Generically one only needs
the lowest-dimensional superfield, in this case $E_\mu{}^\a$, which
has (inverse length) dimension $-\fr2$. All
other superfields will be related to it, and it will contain all the
physical component fields. The method for eliminating
other superfields as independent degrees of freedom is by using {\it
conventional constraints}. They are of two types: those eliminating
the spin connection and those eliminating (part of) the vielbein. I
will not describe the transformations used in order to implement the
conventional constraints; a detailed account can be found in
refs. [\ConvConstr, \CGNN]. The transformations are such that the
transformed fields satisfy the Bianchi identities if the original ones do.

Conventional constraint should be implemented at the level of ``field
strengths'' --- in this case on the torsion. Systematically applying
the associated transformations, it turns out that the torsion can
always be brought to the form
$$
T_{\a\b}{}^c=2\g^c_{\a\b}+\fr2U^c{}_{e_1e_2}\g^{e_1e_2}_{\a\b}
+\fr{5!}V^c{}_{e_1\ldots e_5}\g^{e_1\ldots e_5}_{\a\b}\eqn
$$
\vskip-.5cm\hskip4cm$\uparrow$\hfill\break
\vskip-.5cm\hskip3.5cm{\eightrm standard}

\Yvcentermath1
\null\vskip-12\parskip
\hskip5.2cm$\yng(2,1)$\hskip1.9cm$\yng(2,1,1,1,1)$
\null\vskip4\parskip

\noindent The tensor superfields $U$ and $V$ are all that is left in the torsion
at dimension 0. The Young tableaux indicate the irreducible $so(1,10)$
modules
with this symmetry, which in Dynkin notation will be labelled
$(11000)$ and $(10002)$, respectively. Sofar, the fields remain
off-shell.
It is known that demanding $U=V=0$, \ie, taking the torsion at
dimension 0 to have the ``standard'' form of eq. (\Torsion), implies
the equations of motion. Demanding $U=V=0$ is a {\it physical
constraint}, as opposed to a conventional one. There is no guarantee
that such a constraint does not interfere with the Bianchi identities,
these being integrability conditions on the torsion. Indeed one finds,
by systematically solving the torsion Bianchi identity, that the
equations of motion are forced on the component fields.

All physical fields are, as mentioned, contained in the
supergeometry. For example, the 4-form field strength is found at
dimension 1 as
$$
T_{a\b}{}^\g\varpropto H_{ae_1e_2e_3}(\g^{e_1e_2e_3}){}_\b{}^\g
-\fr8H^{e_1e_2e_3e_4}(\g_{ae_1e_2e_3e_4}){}_\b{}^\g\eqn
$$
See \eg\ ref. [\CGNN] for details. Of course, the vielbein can
contain the 3-form $C$, which is not gauge invariant, only through its
field strength.

There is also a closed superspace 4-form, which contains the bosonic,
physical, one.
The construction of the super-4-form relies on supergeometric data
(the torsion), so this is not an independent construction. However,
$C_{\a\b\g}$ contains the entire linearised supermultiplet, and the
linearised equations of motion are obtained by demanding that the
irreducible modules

\Yvcentermath1
\null\vskip-2\parskip
\hskip2cm
$\yng(2,2)\quad\oplus\quad\yng(2,2,1,1,1)\quad\oplus\quad\yng(2,2,2,2,2)$
\vskip\parskip

\noindent in $H_{\a\b\g\d}$ vanish (the rest are conventional
constraints).

We note that the interesting modules both in $T$ and $H$ are ones
containing columns with 2 and 5 boxes. This is of course no
coincidence. They come from pairs of fermionic indices on the
components in fermionic directions of
superspace forms: $T_{\a\b}{}^a$ and $H_{\a\b\g\d}$. Fermionic form
indices are symmetrised, and the symmetric product of two spinors in
$D=11$ contains a 1-form (vector), a 2-form and a 5-form. Roughly
speaking, the vector part goes away by the conventional constraints,
and the rest remains.

To summarise, the physical fields and equations of motion reside in superfields 
\vskip\parskip
$$
\matrix{E_\a{}^a\,:&\young(a)_\a&\quad\hbox{or}\quad
        &C_{\a\b\g}\,:&\yng(1,1)_\a\oplus\yng(1,1,1,1,1)_\a\cr
\downarrow&&&\downarrow&\cr
T_{\a\b}{}^a\,:&\young(\hfil a,\hfil)\oplus\young(\hfil a,\hfil,\hfil,\hfil,\hfil)&&H_{\a\b\g\d}\,:
&\yng(2,2)\oplus\yng(2,2,1,1,1)\oplus\yng(2,2,2,2,2)\cr}\eqn
$$
\vskip\parskip
\noindent In order to use this information to extract a supersymmetric
action principle, one needs an action containing the upper
superfields, whose 
equations of motion contain the lower ones.
The operation of going from fields to equations of motion looks like
an exterior derivative in a fermionic direction. It indeed is, but in
addition a projection has been performed, where 1-form parts have been
projected away. This will be the r\^ole of the pure spinor.

Remember that a pure spinor $\l$ is defined by
$$
(\l\g^a\l)=0\komma\eqn
$$
so that the non-vanishing bilinears in $\l$ are
$(\l\g^{ab}\l)$ and $(\l\g^{abcde}\l)$. (The precise statement is
particular to $D=11$. In $D=10$, the remaining bilinear is a self-dual
5-form, $(\l\g^{abcde}\l)$.) Now, let us replace the fermionic frame
form $E^\a$ by $\l^\a$. For a $p$-form $\omega$ pointing in the
fermionic directions this simply means replacing 
$$
\omega=\fr{p!}E^{\a_1}\wedge\ldots\wedge
E^{\a_p}\omega_{\a_p\ldots\a_1}\eqn
$$ 
by
$\fr{p!}\l^{\a_p}\ldots\l^{\a_1}\omega_{\a_1\ldots\a_p}$.
In ordinary superspace, taking an exterior derivative means mixing
components with bosonic indices into the result, due to the presence
of torsion:
$$
(d\omega)_{\a_1\ldots\a_{p+1}}=(p+1)D_{(\a_1}\omega_{\a_2\ldots\a_{p+1})}
+{p+2\choose 2}T_{(\a_1\a_2]}{}^a\omega_{|a|\a_3\ldots\a_{p+1})}\komma\eqn
$$ 
where $T_{\a\b}{}^a=2\g^a_{\a\b}$. It is not consistent to treat the
fermionic directions only. However, the second term is projected away
by the pure spinor constraint. So, the projection on certain modules
performed by 
replacing the vielbein by the pure spinor allows for a consistent
treatment of the components along fermionic directions alone.

In this vein, a pure spinor superfield $\Psi(x,\theta,\l)$, with an
expansion
$$
\Psi(x,\theta,\l)=\psi(x,\theta)+\l^\a\psi_\a(x,\theta)
+\fr2\l^\a\l^\b\psi_{\a\b}(x,\theta)+\ldots\eqn
$$
provides a way of dealing with fermionic forms (of arbitrary degree)
in a consistent manner. We will now make the correspondence between the
supergravity vielbein and 3-form and this procedure more precise.

A scalar field $\Psi(x,\theta,\l)$, when expanded in a power series in
$\l$, contains
$$
{}_{1\rightarrow\a\rightarrow\left(\yng(1,1)\oplus\yng(1,1,1,1,1)\,\right)
\rightarrow\left(\yng(1,1)_\a\oplus\yng(1,1,1,1,1)_\a\right)
\rightarrow\left(\yng(2,2)\oplus\yng(2,2,1,1,1)\oplus\yng(2,2,2,2,2)\,\right)
\rightarrow\ldots}\eqn
$$

\noindent We recognise the modules of $C_{\a\b\g}$ and of the equations of
motion.
The cohomology of $Q$, as defined in eq. (\BRSToperator) 
gives the linearised equations of motion!
A completely analogous statement holds for a field $\Phi^a$ and the linearised
supergeometry. In that case $\Phi^a$ enjoys the extra gauge symmetry
$\Phi^a\approx\Phi^a+(\l\g^a\varrho)$ (which can also be understood using
transformations corresponding to
conventional constraints) [\SpinorialCohomology].

This makes it clear how conventional superspace in a natural way leads
to pure spinors. Both the fields and the modules implying the
equations of motion can be interpreted as sitting in a pure spinor
superfield (actually, both come in the same field). This opens for the
possibility of going off-shell for such a field. The linearised equations of
motion will be encoded as $Q\Psi=0$ or $Q\Phi^a=0$. 

Before turning to examining the implications of this, I would like to
say some words about the pure spinor space. The pure spinor constraint
only has solutions for complex $\l$, and the solution space turns out
to be 23-dimensional. 9 out of the 11 constraints on the
32-dimensional spinor are independent.

\vskip2\parskip
\epsffile{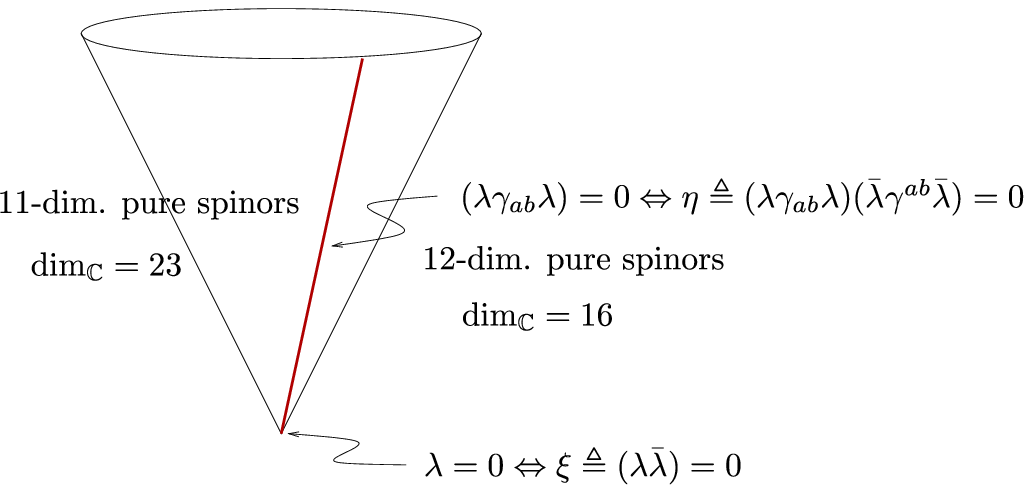}

\noindent There is a special 16-dimensional 
subspace of the pure spinor c\^one where not only
$(\l\g^a\l)$, but also $(\l\g^{ab}\l)$ vanishes. This is the space of
12-dimensional pure spinors. Here is a difference from $D=10$ where
any monomial in $\l$ consists of one irreducible module. While the
only singular point in $D=10$ pure spinor space is the tip of the
c\^one, $\l=0$, there is a singular subspace in $D=11$. This will be
relevant later, when we consider operators on the pure spinor space.

The pure spinor superfields considered earlier are holomorphic in the
complex variables $\l^\a$. This (and other issues) raises the question
of how integration with respect to $\l$ should be performed.
By looking at the cohomology of the BRST operator $Q$, we will get a
hint.
We have already seen that the cohomology (at $\l^3$ in the scalar
field $\Psi$, corresponding to the superfield $C_{\a\b\c}$, and at
$\l$ in $\Phi^a$, corresponding to the linearised field $E_\mu{}^a$)
contains the physical fields. There is clearly a gauge
invariance, \eg\ $\delta\Psi=Q\Lambda$. A careful examination shows
that there is a cohomology in $\Psi$ at $\l^2$ containing gauge
transformations (not only tensor gauge transformations, but also
diffeomorphisms and local supersymmetry). But since $\l$ has ``wrong''
statistics, this cohomology in $\Psi$ will have opposite statistics
compared to gauge parameters. These are ghost fields. Indeed, it turns
out that the cohomology encodes also the reducibility of the tensor
gauge transformations/ghosts, all the way down to the scalar
ghost-for-ghost-for-ghost, which sits as the $\l$- and
$\theta$-independent part of $\Psi$. Corresponding statements are true
of $\Phi^a$, but the gauge transformations encoded are only
diffeomorphisms and local supersymmetry. 

Let us for a moment specialise on the zero-mode cohomology, \ie, the
cohomology of an $x$-independent field $\Psi(\theta,\l)$. Why? We have
argued that $Q\Psi=0$ is the condition that enforces the equations of
motion, which are some differential equations with respect to $x$. For
zero-modes, they are automatically satisfied, and the cohomology
problem turns into a purely algebraic problem. It can be solved by
hand, or with computer assistance.
A little thinking also tells us that if there are equations of
motion imposed by cohomology on the fields at $\l^p$, these must be
represented in the zero-mode cohomology at $\l^{p+1}$. 
In addition to the zero-mode cohomology giving physical fields and ghosts, there
is in the cohomology of $\Psi$ a complete ``mirror'' of fields, where
the same (more generically, conjugate, but all $so(11)$ modules are
self-conjugate) modules occur at $\l^p$ and $\l^{7-p}$. The cohomology
at $\l^4$ has the right properties to represent field equations or
currents (as we have already argued earlier). The ``wrong'' statistics
again forces an interpretation on us: they are antifields, in the
Batalin--Vilkovisky (BV) [\BatalinVilkovisky] sense. All fields and
ghosts and their respective antifields naturally occur as cohomology.
A complete table of the zero-mode cohomology in $\Psi$ is given
below. The modules are given with their Dynkin label.

\epsfxsize=9cm
\epsffile{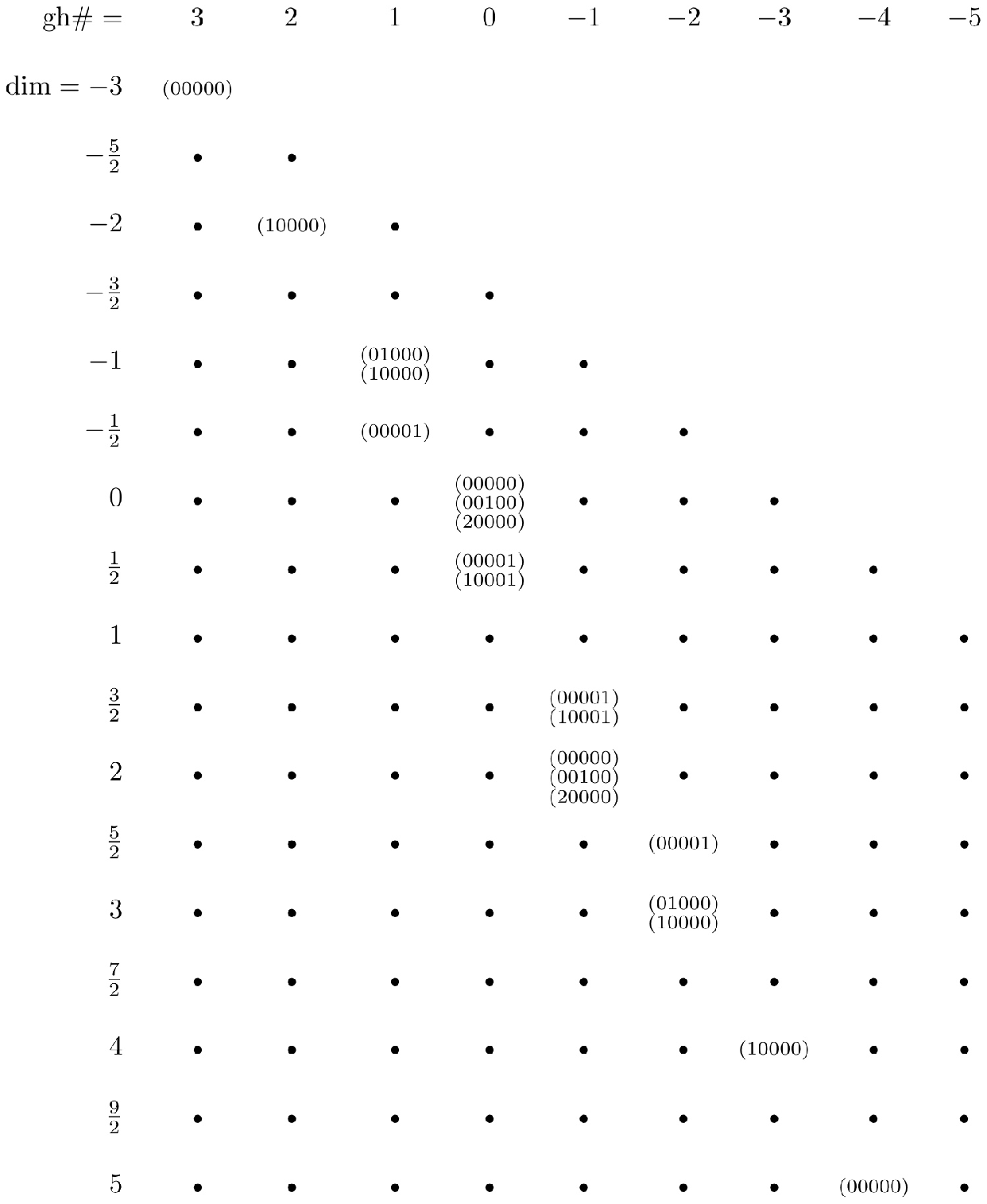}

In the table, ghost numbers and dimensions have been assigned by
demanding that those of the physical fields are correct. I would like
to point at the ``highest'' cohomology, corresponding to the antifield
for the ghost-for-ghost-for-ghost. This component of $\Psi$, itself a
field with dimension $-3$ and ghost number 3, has dimension 5 and
ghost number $-4$. Suppose we try to define integration as a kind of
residue, by taking the component of an integrand in this
cohomology. Such an integration would have dimension $-8$ and ghost
number $-7$. Consider a linearised action of the type
$$
\kappa^2S_0\sim\int\Psi Q\Psi\punkt\Eqn\LinearisedAction
$$  
Together with integration over theta, the total dimension and
ghost number of
the component Lagrangian would be 
$$
\eqalign{
\hbox{dim}(\kappa^2L_0)&=2\times(-3)-8+\fr2\times32=2\komma\cr
\hbox{gh\#}(\kappa^2L_0)&=2\times3+1-7=0\punkt\cr
}\eqn
$$
This matches perfectly for a component Lagrangian. 

One may therefore think that this solves the problem of finding a
linearised action for the pure spinor superfield $\Psi$, whose
equation of motion is $Q\Psi=0$ and reproduces the linearised
supergravity multiplet, at least around flat space. This is not yet
the case, but it is a big step on the way. The remaining problem lies
in the observation that the integration is singular. A ``residue''
does not provide a non-singular measure when the power expansion is
limited from one side. It is a bit like trying to define a residue for
polynomials. This difficulty was resolved by the introduction of
non-minimal variables [\BerkovitsNonMinimal]. 
In addition to $\l$, one also considers
the pure spinor $\bar\l$ and a fermionic spinor $r^\a$ which is pure
relative to $\bar\lambda$, $(\bar\l\g^ar)=0$. The BRST operator is
changed into 
$$
Q=\l^\a D_\a+r_\a{\*\over\*\bar\l_\a}\eqn
$$
This does not affect the cohomology. I will not go further into details
about non-minimal variables and integration in this lecture.

I would now like to discuss the question of interactions. But first, a
few words on the two fields, $\Psi$ and $\Phi^a$. Each of them is
capable of completely describing the linearised supergravity
multiplet. An important difference is that while $\Psi$ contains the
``naked'' 3-form potential and the associated ghosts, $\Phi^a$ does
not. It only contains the 3-form through its 4-form field strength
$H=dC$. A further observation is that the cohomology of $\Phi^a$ (not
presented in detail above), although having a kind of ``mirror
symmetry'', does not show a symmetry between fields and
antifields. When one goes beyond equations of motion, the cohomology
looks ``too big''. Neither does this cohomology possess a singlet that
can be related to a measure. 

Since the component action contains a
Chern--Simons term $\sim C\wedge H\wedge H$, it can never be
constructed from $\Phi^a$ alone. We must think of $\Psi$ as the
fundamental field and $\Phi^a$ as a derived field. It therefore seems
likely that there is some operator $R^a$, such that
$\Phi^a=R^a\Psi$. Since cohomology should map to cohomology, $R^a$
itself should commute with $Q$ (modulo gauge transformation 
$\delta\Phi^a=(\l\g^a\varrho$). 
It is indeed possible to construct such an operator, with the correct
quantum numbers. Here, I will not give the full form.
$$
R^a=\eta^{-1}(\lb\g^{ab}\lb)\*_b+\ldots\komma\eqn
$$
where the ellipsis denotes terms with $r$. $\eta$ is the invariant
vanishing on the 16-dimensional subspace of 12-dimensional pure
spinors. This is again a difference from $D=10$, where operators with
negative ghost number typically diverge only at $\l=0$. 
The operator $R^a$ turns out to provide key input
for the construction of interaction terms.

It turns out to be very fruitful to play with the fields $\Psi$ and
$\Phi^a$, and ask for possible 3-point couplings matching the counting
of dimension an ghost number (no dimensionful constants should
be included, unless one looks for higher-derivative interactions
[\CGNN]). The extra gauge invariance for $\Phi^a$ can be taken care of
by demanding that it always sits in a combination
$(\l\g_{ab}\l)\Phi^b$. Then the Fierz identity
$(\l\g_{ab}\l)(\g^b\l)^\a=0$, holding for a pure spinor, assures gauge
invariance. Such a factor can also help to contract the vector indices
on two (fermionic) $\Phi$'s. Simple counting shows that the
combination 
$$
S_1\sim\int\Psi(\l\g_{ab}\l)\Phi^a\Phi^b
=\int(\l\g_{ab}\l)\Psi R^a\Psi R^b\Psi\eqn
$$
is the only gauge invariant combination of $\Psi$'s and $\Phi$'s, without extra
operators, that has the correct dimension and ghost number.

Could this be a good 3-point coupling?
What are the principles in deciding which interaction terms are allowed?
In order to answer these questions, we need to talk a little more about
the BV formalism\foot\star{In principle, one analyse interactions in
terms of gauge invariance. But since both the action and the gauge
transformations may get modifications, the BV framework turns out to
be much more efficient, in that it deals with both issues at
once.}\hskip-3pt. A good review, departing from classical field
theory, 
is provided in ref. [\FusterBVReview]. 

The BV formalism, in general, builds on a ``doubling'' of all fields,
physical ones as well as ghosts, with their corresponding
antifields, of opposite statistics. 
A fundamental structure, similar to a Poisson bracket, is provided by
the antibracket, which in a component formalism is defined as
$$
(A,B)=\int[dx]
\left(A{\larrowover\d\over\d\phi^A(x)}{\rarrowover\d\over\d\phi^\star_A(x)}B
-A{\larrowover\d\over\d\phi^\star_A(x)}{\rarrowover\d\over\d\phi^A(x)}B\right)
\punkt\eqn
$$ 
Here, $\phi^A$ denote fields (including ghosts) and $\phi^\star_A$
antifields.
The action itself is the generator of ``gauge transformations'',
generated as $\delta X=(S,X)$, where $(\cdot,\cdot)$ is the
antibracket. 
The governing equation generalising $Q^2=0$ is the BV master equation
[\BatalinVilkovisky]
$$
(S,S)=0\komma\eqn
$$
and this is the only consistency check needed when introducing interactions.

BRST cohomology is an inherently linear concept, and the BV formalism
is the appropriate way to generalise it to non-linear (interacting)
theories.
Since we already know the the BRST cohomology of a pure spinor
superfields provides both fields and antifields, there is no choice
but to follow the BV procedure.
The difference from a component formulation is that we are dealing
with a single field $\Psi$, encoding all fields and antifields.
For the pure spinor superfield $\Psi$, the antibracket takes the 
simple form [\PureSG]
$$
(A,B)=\int
A{\larrowover\d\over\d\Psi(Z)}[dZ]{\rarrowover\d\over\d\Psi(Z)}B
\komma\eqn
$$
which I interpret as another sign that we are on the right track (the
integral here is over all variables).

The full BV action for $D=10$ super-Yang--Mills (and its dimensional
reductions) is the Chern--Simons-like action
$$
S=\int[dZ]\Tr\left(\fr2\Psi Q\Psi+\fr3\Psi^3\right)\eqn
$$
(implicit in refs. 
[\BerkovitsParticle,\BerkovitsNonMinimal,\CederwallNilssonTsimpisI]).
Note that there is only a 3-point coupling; the quartic interaction
arises on elimination of ``auxiliary fields'', notably the lowest
component in the superfield $A_\a(x,\theta)$.

An analogous formulation exists for the Bagger--Lambert--Gustavsson
and Aharony--Bergman--Jafferis--Maldacena models in $D=3$. 
The
simplification there is even more radical: The component actions
contain 6-point couplings, but the pure spinor superfield actions only
have minimal coupling (\ie, 3-point interactions) 
[\CederwallBLG,\CederwallABJM].

But I would like to turn back to supergravity.
The fact that the operator $R^a$ commutes with $Q$ (modulo gauge
transformations) ensures that the
interaction term proposed above
$$
S_1\varpropto\int[dZ](\l\g_{ab}\l)\Psi R^a\Psi R^b\Psi\eqn
$$
is a nontrivial deformation respecting the master equation.
The factor $(\l\g_{ab}\l)$

\item{$\bullet$}{ensures that dimension and ghost number are correct,}

\item{$\bullet$}{guarantees the invariance under 
           $\Phi^a\approx\Phi^a+(\l\g^a\varrho)$,}

\item{$\bullet$}{makes possible a contraction of $\Phi^a$'s.}

\noindent Some terms have been checked explicitly 
(Chern--Simons term, coupling of
diffeomorphism ghosts encoding the algebra of vector fields), 
so it is clear that this gives the 3-point
couplings of $D=11$ supergravity.

One may expect that an expansion around flat space would be
non-polynomial. This is however not the case. Checking the master
equation to higher order in the field involves commutators of
$R^a$'s. The $R^a$'s don't commute, but ``almost''.
$$ 
\fr2(\l\g_{ab}\l)[R^a,R^b]=\Fr32\{Q,T\}\eqn
$$
where $T=8\eta^{-3}(\lb\g^{ab}\lb)(\lb r)(rr)(\l\g_{ab}w)$.
The master equation is {\it exactly} satified by
$$
S=\int[dZ]\left[\fr2\Psi Q\Psi+
\fr6(\l\g_{ab}\l)(1-\Fr32T\Psi)\Psi R^a\Psi R^b\Psi\right]\punkt\eqn
$$
Note the similarity of the 3-point coupling 
($\varpropto\Psi\Phi\Phi$)
to the Chern--Simons term
(which it indeed contains).
After a field redefinition $\Psi=(1+\fr2 T\TPsi)\TPsi$:
$$
S
=\int[dZ]\left[\fr2(1+T\TPsi)\TPsi Q\TPsi+
\fr6(\l\g_{ab}\l)\TPsi R^a\TPsi R^b\TPsi\right]\punkt\eqn
$$
I would like to stress that this is quite a remarkable property. It
seems that the elimination of auxiliary fields will reintroduce the
non-polynomial property of the component supergravity.
There is of course a price for this simplicity. The geometric picture
is lost, when the fields are expanded around a background (in our
case, a flat one). Even if the action is exact to all orders, it is
not clear how to find solutions that correspond to exact solutions in
gravity or supergravity.

A few words on gauge fixing. In the BV formalism, it amount to
ordinary gauge fixing of the physical fields, as well as elimination
of antifields.
Covariant gauge fixing (Siegel gauge) amounts to demanding 
$$
b\Psi=0\komma
$$
where $b$ is the composite $b$-ghost, satisfying 
$[Q,b]=\square$.
The propagator then becomes $b\square^{-1}$.
Unlike in component BV formalism, there is no need to introduce
non-minimal fields (antighost, Nakanishi--Lautrup field); they are
contained in $\Psi$ (implicit in ref. [\AisakaBerkovits]).
The $D=11$ $b$-ghost has been constructed [\ABC], and takes the form
$$
b=\fr2\eta^{-1}(\bar\l\g_{ab}\bar\l)(\l\g^{ab}\g^iD)\*_i+\ldots
$$

Some conclusions and problems:

\item{$\bullet$}{The framework described resolves the issue of
classical supersymmetric actions
for maximally supersymmetric theories.}

\item{$\bullet$}{The interaction terms are generically 
much simpler and of lower order
than in a component language; for supergravity to the extent that the action
becomes polynomial.}

\item{$\bullet$}{Presumably, the formalism may be efficient  
for calculating quantum
amplitudes. Need to establish connection to ``superparticle'' prescription.
Finiteness of BLG? Of $N=8$ supergravity?
Regularisation is needed in path integrals, due to negative powers of $\eta$.} 

\item{$\bullet$}{How is U-duality realised?
Models connected to 
generalised geometry, with enlarged structure groups, may possibly
provide generalised models of gravity?}

\item{$\bullet$}{Geometry? 
Background invariance? The polynomial property should be better understood.}

\refout
\end